\documentclass[a4paper,11pt]{article}
\usepackage{jinstpub} 
\usepackage{lineno}

\usepackage{pgfplots}
\usepgfplotslibrary{external}

\subheader{Review}
\proceeding{Topical Workshop on Electronics for Particle Physics - TWEPP2023\\
  Oct 1--6, 2023 \\
  Geremeas, Sardinia, Italy}

\title{Lessons learned while developing the Serenity-S1 ATCA card}

\author[a]{T. Mehner} 
\author[a]{L.E. Ardila-Perez} 
\author[a]{M. Balzer}
\author[b]{G. Fedi}
\author[a]{M. Fuchs} 
\author[b]{A. Howard}
\author[b]{G. Iles}
\author[g]{M. Loutit}
\author[b]{S. Mansbridge}
\author[d]{F. Palla}
\author[b]{D. Parker}
\author[b]{M. Pesaresi}
\author[b]{A. Rose}
\author[b]{M. Saleh}
\author[a]{O. Sander} 
\author[a]{M. Schleicher}
\author[f]{C. Strohman}
\author[a]{D. Tcherniakhovski}
\author[e]{T. Williams}
\author[c]{J. Zhao}

\affiliation[a]{Karlsruhe Institute of Technology, Institute for Data Processing and Electronics,\\Hermann-von-Helmholtz-Platz 1, D-76344 Eggenstein-Leopoldshafen, Germany}
\affiliation[b]{Imperial College London, Physics Department, Blackett Laboratory,\\Prince Consort Rd, London, SW7 2BW, UK}
\affiliation[c]{Chinese Academy of Sciences, Institute of High Energy Physics,\\19B Yuquan Road, Shijingshan District, Beijing, China}
\affiliation[d]{National Institute of Nuclear Physics, Pisa Section,\\56124 Pisa, Italy}
\affiliation[e]{STFC Rutherford Appleton Laboratory, Particle Physics Department,\\Harwell Campus, Didcot, OX11 0QX, UK}
\affiliation[f]{Cornell University, Department of Physics,\\616 Thurston Ave., Ithaca, NY 14853, USA}
\affiliation[g]{University of Bristol, School of Physics,\\Queens Road, Bristol, BS8 1QU, UK}

\emailAdd{torben.mehner@kit.edu}

\abstract{
The Serenity-S1 is a production-optimised Advanced Telecommunications Computing Architecture (ATCA) processing blade based on the AMD Xilinx Virtex Ultrascale+ device. It incorporates many developments from the Serenity-A and Serenity-Z prototype cards and, where possible, adopts solutions being used across CERN. Due to the shortage of components during the recent semiconductor crisis, commonly used components in the prototypes had to be replaced by new ones after qualification.  
In this work, we discuss various improvements to simplify manufacturing, the performance of new components, some of the more difficult aspects of procurement, the performance of production-grade Samtec 25\,Gb/s optical firefly parts, and concerns regarding the rack cooling infrastructure.
}


\arxivnumber{2311.02222} 

\notoc

\begin{document}
\maketitle
\flushbottom

\section{Introduction}

The Serenity-S1 ATCA card is an integral component of the High-Luminosity Large Hadron Collider (HL-LHC) upgrade\,\cite{bortolato_architecture_2023, Ryd_2020}. A considerable number of 721 Serenity-S1 cards will be integrated into various sub-detector systems within the Compact Muon Solenoid (CMS) experiment. As a result, a versatile design is required, featuring optical high-speed data input and output capabilities as well as a large processing FPGA. The Serenity-S1 card evolved from its predecessors, namely the Serenity-A and Serenity-Z\,\cite{Mehner_2022, rose_serenity_2019} prototype cards, which were developed at Karlsruhe Institute of Technology (KIT) and Imperial College London, respectively. These prototype cards are already in use for firmware development and system integration in several CMS sub-detector working groups. The Serenity-S1 is the result of a synergistic partnership within the Serenity consortium, encompassing multiple institutes across numerous countries. This cooperative design process was facilitated by a Git-controlled Altium project, enabling concurrent work on the schematic and sequential work on the layout. Enabling this workflow was a key reason why we shifted to Altium for the development of Serenity-S1. Furthermore, the real-time monitoring of component availability that Altium offers was also used intensively, especially during the semiconductor crisis.

\section{Board Overview}

\begin{figure}[htbp]
    \centering
    \includegraphics[width=.3\linewidth]{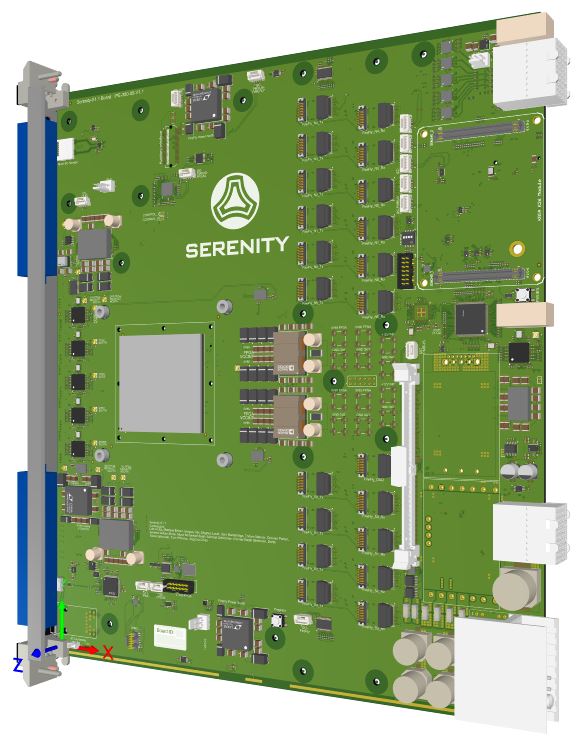}
    \caption{3D render of the Serenity-S1 ATCA board.}
    \label{fig:serenity-s1}
\end{figure}

The board design as seen in \autoref{fig:serenity-s1} incorporates segmentation into two distinct sections, namely the payload area towards the front panel and the service area towards the backplane. This segmentation was motivated by an initial plan to develop both a single and a double FPGA version. At this time, only the single FPGA version is foreseen within CMS. The division into payload area and service area is nevertheless advantageous, as the service area, which is dedicated to the slow control of the board, can be reused for future ATCA developments. Within the service area, an AMD Xilinx Kria K26 System-on-Module (SoM) is running a Linux operating system to manage the slow control tasks\,\cite{Mehner_2022, Fuchs_2023}. This section also houses the required power supplies, an Ethernet PHY, an SD card slot, SSD storage, and clocking resources tailored for the Kria module. The second portion of the service area contains what is needed to meet the specific requirements of the ATCA standard. This includes an Intelligent Platform Management Controller (IPMC) named OpenIPMC in the DIMM form-factor\,\cite{Calligaris_2022}, Zone-2 backplane connections, the necessary input power modules designed to harness the dual -48\,V shelf power source, and an Ethernet switch that allows connecting up to seven endpoints. The payload area is responsible for processing data with a high throughput and low latency. The main processing unit in this domain is an AMD Xilinx Virtex Ultrascale+ VU13P FPGA. Data transmission is handled by ten pairs of 12-channel unidirectional FireFly connectors and one additional 4-channel bidirectional connector. Furthermore, the payload area houses a synchronous and asynchronous clock tree, providing reference clocks to all multi-gigabit transceivers (MGTs) of the FPGA.

\section{Component Selection Challenges}

\begin{figure}[htbp]
    \centering
    \begin{tikzpicture}
        \begin{axis}[
            xlabel={Frequency Offset [Hz]},
            ylabel={Phase Noise [dBc/Hz]},
            legend pos=north east,
            ymajorgrids=true,
            xmajorgrids=true,
            xmode=log,
            width=\linewidth,
            height=0.5\linewidth]
        
            \addplot table [x=A, y=B, col sep=comma, mark=none] {ldos.csv};
            \addlegendentry{LDO: jitter = 997.758\,fs}

            \addplot table [x=A, y=B, col sep=comma, mark=none] {noldos.csv};
            \addlegendentry{No LDOs: jitter = 838.522\,fs}

            \addplot[x=A, y=B, mark=square] coordinates{
                (10000, -112)
                (100000, -128)
                (1000000, -145) };
            \addlegendentry{Virtex Ultrascale+ GTY Phase Noise Mask}
        \end{axis}
    \end{tikzpicture}
    \caption{Phase noise measurements for ZL30274 with and without LDO (1\,Hz to 10\,MHz).}
    \label{fig:jitterresults}
\end{figure}
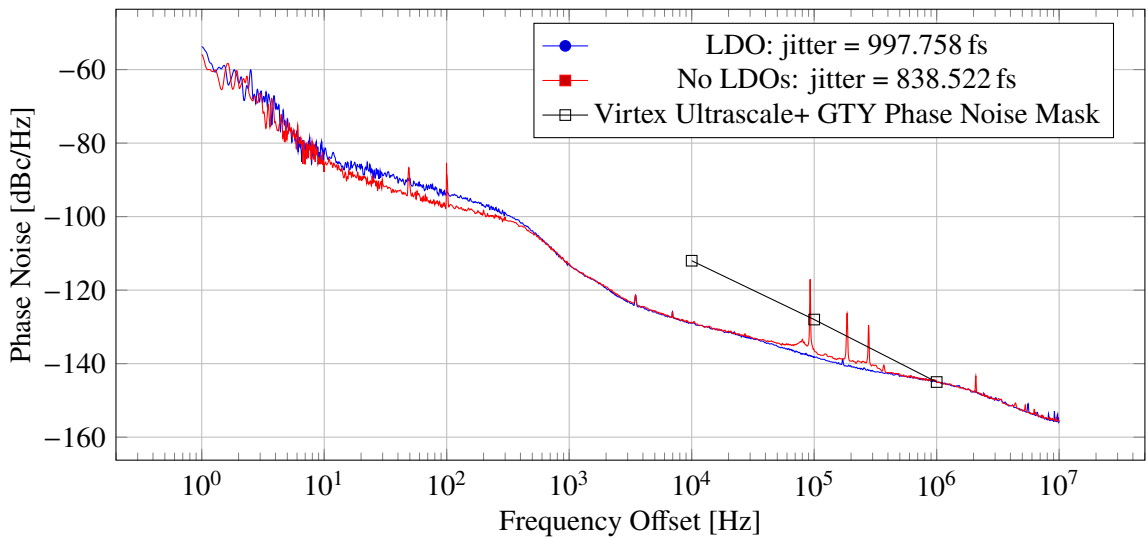

The initial choice for the clock distribution network, the Skyworks Si5395A zero-delay jitter cleaning phase-locked loop (PLL), became unavailable without any indication of its future availability. Subsequently, a suitable alternative was identified, the Microchip ZL30274, which features two internal PLLs and similar phase-noise performance. An evaluation card was developed and sent to the EP-ESE group at CERN for assessment of compatibility with the CMS requirements. Extensive testing, including measurements with and without the on-board low-dropout regulators to filter the switched-mode power supplies, shows that the ZL30274 met the demands of the Virtex Ultrascale+ FPGA \cite{amd_ds923} and the $\leq$10\,ps jitter specification of CMS in general \cite{Collaboration:2759072} (see \autoref{fig:jitterresults}). As a consequence, the ZL30274 emerged as a viable replacement for the Si5395A in the design.

Similar concerns arose regarding the core power supply, which is delivering 170\,A at 0.85\,V to the FPGA. While the initial choice involved two parallel LTM4700 power modules, their limited availability and a subsequent 70\,\% price increase motivated an exploration of alternatives. Due to the availability relaxing, the LTM4700 was kept in the design, but the option to use a multi-phase power supply that can be built using industry-standard components was added. The multi-phase supply can be mounted using Würth Electronic REDCUBE terminals that can transmit 50\,A each. As soon as the newly designed multi-phase power supply has been successfully tested, it can replace the LTM4700 modules, which currently cost \$\,300 while the multi-phase power supply will be priced at \$\,40. This underscores the importance of planning for alternative solutions when component availability is uncertain.

Recently, the production of the standardised power connector for ATCA cards has been halted by multiple manufacturers. While Positronic and Conec still produce this connector, there is a worrying trend of diminishing support for ATCA connectors. In response, we adopted a multi-vendor-compliant approach by adapting our footprints and accounting for alignment pin positions. Additionally, the total number of ATCA connectors required for Serenity-S1 boards in CMS, plus a reserve of 15\,\%, has already been acquired.

\section{PCB Design}

\begin{figure}[htbp]
    \centering
    \includegraphics[height=.4\textwidth]{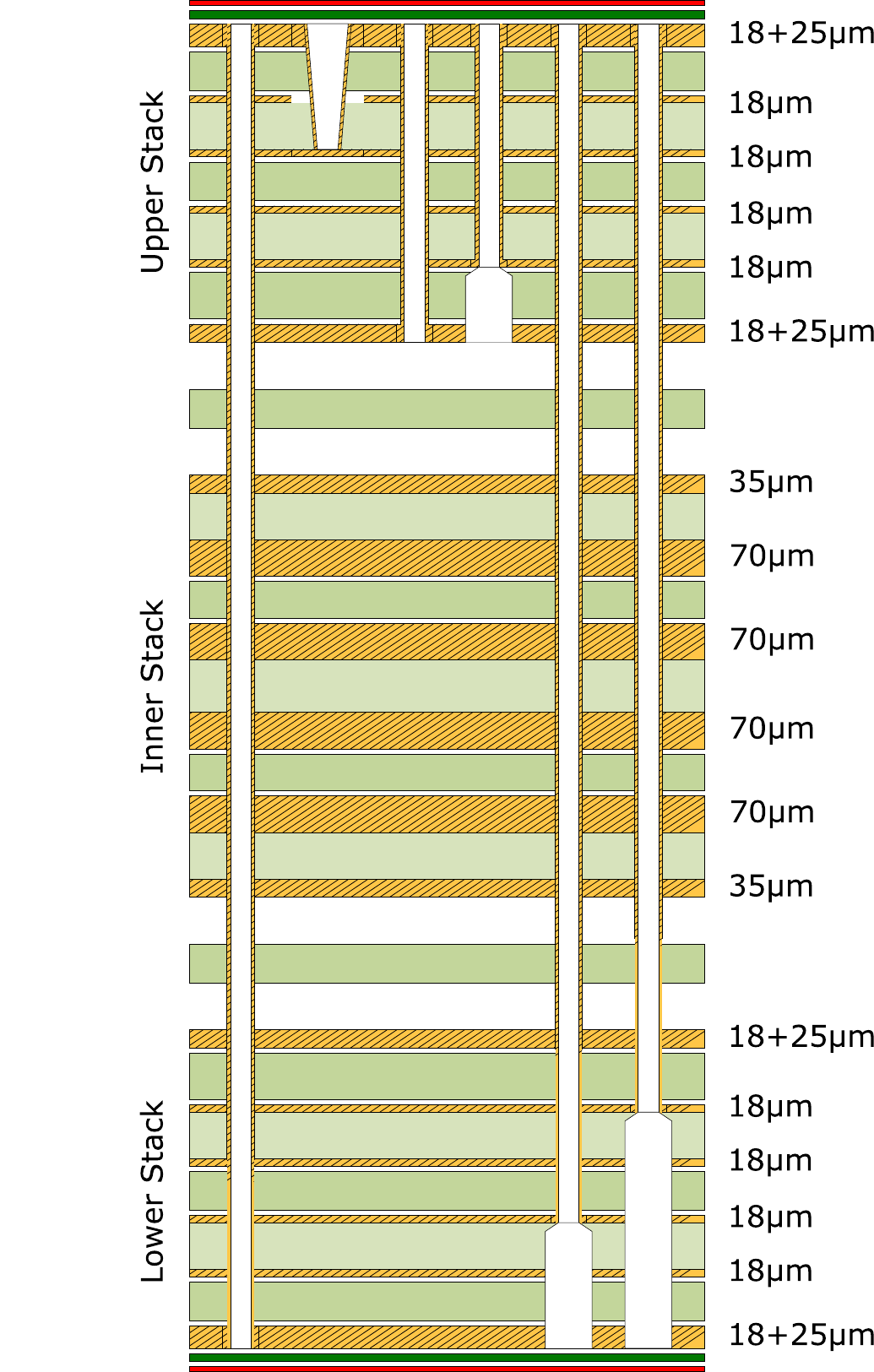}
    \qquad \qquad
    \includegraphics[height=.4\textwidth]{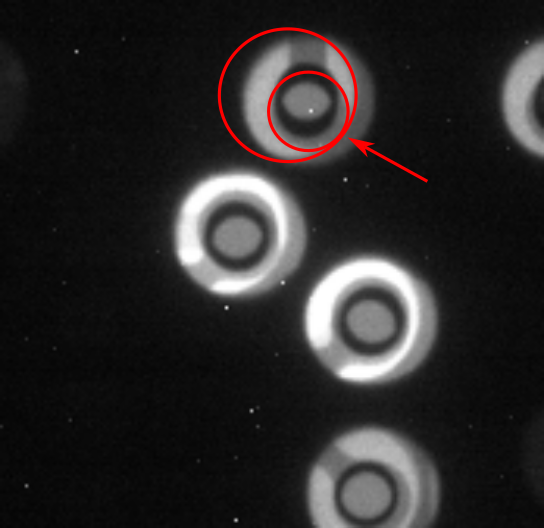}
    \caption{Serenity-S1 layer stack (left) and x-ray picture showing a short of a via pad (small circle) with the copper flood (large circle) due to center stack expansion (right).}
    \label{fig:xrayshort}
\end{figure}

The PCB design presented three significant challenges. First, the 124 high-speed links from the FPGA to the FireFly connectors require a minimum of three impedance-controlled, ground-shielded layers that are accessible from the top through stub-less vias. Second, to limit thermal losses due to high currents in the power planes, the copper thickness was increased from the standard 18\,µm used for the signal layers to 70\,µm. To further reduce power distribution losses, additional layers were used where feasible. Lastly, the stipulation by the experiment to use halogen-free materials posed a challenge due to the limited experience of PCB manufacturers with high-speed halogen-free materials, prompting close collaboration with manufacturers to define a suitable layer stack. The result is shown in \autoref{fig:xrayshort} (left).

The resulting PCB contains three sub-stacks to satisfy layer stack symmetry while incorporating a copper-heavy inner stack and two impedance-controlled outer stacks. The outer two sub-stacks include four impedance-controlled layers accessible from the top via back-drilling and microvias. The inner stack conducts high currents across four 70\,µm copper layers, with two allocated for core supply current. To save cost, the PCB supplier initially intended to use a low-speed material for the inner layer but encountered complications due to mismatched expansion parameters between high-speed and low-speed materials. This resulted in shorts around the edge of the PCB, as seen on the right side of \autoref{fig:xrayshort}. 

\section{Cooling System Design and Testing}

Cooling the card requires a custom heatsink solution to maintain a low profile, as the ATCA standard only allows components to elevate to a maximum of 21.33\,mm above the PCB. Additionally, the ATCA crates, which house the cards, require heat extraction to achieve the desired operating temperatures. In the CMS setup, each rack accommodates two ATCA crates, cooled by internal air loops, and three water-cooled heat exchangers to remove the heat. Given the maximum allowed power usage of 10\,kW in a full rack, an equivalent amount of heat must be removed. Maintaining FireFly optical devices below 50\,°C is crucial for their longevity over the targeted 10 years of operation. This results in the boundary conditions that the cooling system must meet. To assess cooling efficiency, measurements were conducted using heater cards at the tracker integration facility (TIF) and at the CMS service cavern USC55\,\cite{Fedi_2023}. With the rack operating at maximum power and the heat exchangers supplied with water at 16\,°C, it was initially only possible to cool the circulating air down to 30\,°C. Increasing the rack water flow from 18\,L/min to 30\,L/min improved this to 26\,°C. Larger-diameter pipes to the rack should yield a rack water flow of 34\,L/min and a resulting air temperature of 25\,°C. This should provide adequate headroom to cool the optical devices. Further improvements are unlikely because the water temperature cannot be lowered for legacy reasons, and it is also not possible to use higher pressure to increase water flow.

\section{FireFly Long Trace Performance}

\begin{figure}[htbp]
    \centering
    \includegraphics[width=\textwidth]{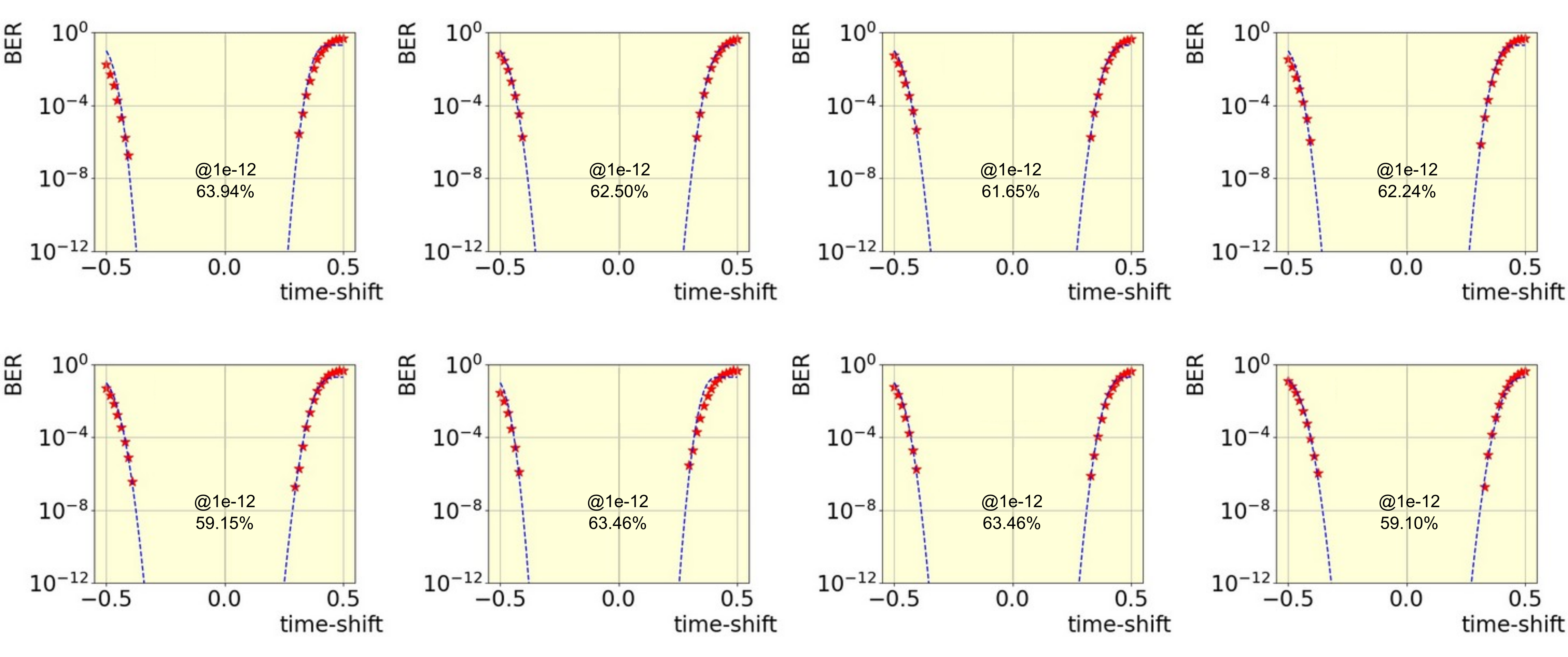}
    \caption{Bathtub plot of Serenity-A long trace (260\,mm).}
    \label{fig:bathtub}
\end{figure}

To safeguard the FireFly transceivers from the FPGA-generated heat, they are placed within separate air passages. However, this arrangement necessitates extended traces running from the FPGA to the FireFly connectors. On the Serenity-S1, these traces span 270\,mm, prompting concerns regarding the ability to achieve 25\,Gb/s links with a low bit error rate of $10^{-12}$. Benefiting from the similar structure in Serenity-A, these long traces could be assessed in advance. A tool was developed to scan the optical performance for different values of the parameters Tx pre-emphasis, Tx post-emphasis, Tx main driver, and Rx Equalizer. The tool allows the evaluation of the operational regions of a device in parameter space and determines if all links work without error. The bathtub plots resulting from these evaluations (see \autoref{fig:bathtub}) demonstrated an opening ranging between 59.10\% and 63.94\%, indicating that the long traces on the Serenity-S1 will likely be functional. This underscores the value of functional evaluation boards for testing purposes. The tests were conducted using prototype FireFly devices, with the first production devices anticipated at the end of October 2023.

\section{Testing and Production Plans}

The first set of 12 boards will imminently arrive from the PCB manufacturer at KIT and undergo assembly there. These initial boards will not have an FPGA populated to facilitate preliminary tests on the power supplies. Subsequently, once the FPGA is socketed or soldered onto the boards, further testing will be conducted before dispatching them to CERN and Imperial College for temperature cycling tests, optical assessments, and performance characterisations of the high-speed lanes.

The full-scale production is scheduled to commence in October 2024, following the tenders starting in February. A pre-series consisting of 50 boards will be the first to be manufactured, with the remaining boards planned for 2025. Separate tenders are being issued to PCB manufacturers and assemblers, with critical components provided to support pre-production, including all ATCA connectors, which will be provided for the entire production run. The high cost of the components necessitates a high yield. In order to achieve this, the board is being designed for ease of manufacture. Very small components are avoided where possible. Dedicated test points will allow in-circuit testing of power nets before power is applied. All power supplies are under I2C control, providing diagnostics across the board. Lastly, large through-hole components will not directly connect to power planes so that they are thermally isolated, allowing safe reflow and rework.

\section{Conclusion}

In conclusion, the Serenity-S1 ATCA card developed for the HL-LHC upgrade demonstrates adaptability in overcoming component selection challenges, complex printed circuit board construction, and cooling system complexities. The evaluation of alternative components and multi-vendor-compliant approaches enhances the resilience of the project. Additionally, by engaging in collaborative efforts with multiple PCB manufacturers, PCB production becomes more dependable and consistent. The construction of prototype cards facilitates early hardware and firmware evaluation, which is crucial for system integration studies at every sub-detector system. The Serenity-S1 project represents a collaborative and innovative approach, well-prepared to meet the demands of the HL-LHC upgrade.

\acknowledgments
The authors would like to acknowledge the support from the CERN EP-ESE group and specially Philippa Hazell and Sophie Baron for the measurements of the phase-noise of the ZL30274 jitter cleaner. Marvin Fuchs acknowledges the support by the Doctoral School ``Karlsruhe School of Elementary and Astroparticle Physics: Science and Technology''.


\bibliographystyle{JHEP}
\bibliography{biblio.bib}






\end{document}